\documentclass[aps,prb,twocolumn,showpacs,superscriptaddress,groupedaddress,floatfix]{revtex4-1}
\usepackage{latexsym}
\usepackage{graphicx}
\usepackage{dcolumn}
\usepackage{bm}
\usepackage{amssymb}
\usepackage{amsmath}
\usepackage[space]{grffile}
\usepackage{color}

\begin{document}

\title{Quench dynamics of superconducting fluctuations and optical conductivity in a disordered system}
\date{\today}
\begin{abstract}
There has been significant interest in the generation of very short-lived superconducting states in solid-state films. Here we consider the role of
non-equilibrium superconducting fluctuations in such systems, generated by an interaction quench, considering the limit of large static disorder.
In particular, we argue that because of critical slowing down, the regime of the fluctuation dominated \emph{normal state}
is more important than might be na\"{i}vely thought. We show how such a state might appear in the optical conductivity, and give the appropriate non-equilibrium
generalization of the Azlamazov-Larkin and Maki-Thompson fluctuation corrections to the optical conductivity. For a quench to
the superconducting critical point, we show that the fluctuation corrections lead to power-law aging behavior in the optical
conductivity. The power-law aging from Azlamazov-Larkin and Maki-Thompson terms are qualitatively different, despite the effect of the
two being similar in dc measurements in thermal equilibrium. These signatures provide a road-map
for experiments to identify the role of superconducting fluctuations in a transient state.
\end{abstract}
\author{Yonah Lemonik}
\author{Aditi Mitra}

\affiliation{Center for Quantum Phenomena, Department of Physics, New York University, 726 Broadway, New York, NY, 10003, USA}
\maketitle

\section{Introduction}
Ultra-fast spectroscopy can not only probe dynamics of an electron system on time scales shorter than the
thermalization time~\cite{Fausti11,Smallwood14,Averitt16}, but may even induce
non-equilibrium phase transitions through the pump laser~\cite{Mitrano15,Knap16,Kennes17,Sentef17,Sentef17b}. Thus
a particular topic of interest is studying the out-of-equilibrium behavior of systems near a critical point.
Here we address the dynamics of a thin metallic film where the distance to the superconducting critical point varies
rapidly in time, but the system remains in the normal state.  We discuss the impure limit
$T\tau/\hbar \ll 1$, where $T$ is the temperature and $\tau$ the elastic impurity scattering time
- the clean limit is discussed elsewhere~\cite{Lemonik17,Lemonik17b}.

For a system in thermal equilibrium at temperature $T$, and in the vicinity of a superconducting critical point, an important quantity
is the detuning from the critical point $\epsilon \equiv (T-T_c)/T$. Here $T_c(U)$ is the critical temperature for an interaction strength $U$.
We may imagine
that the pump field is capable of temporally controlling the detuning $\epsilon$ via a temporal control of the interaction $U$.
In this paper we study quench dynamics for some representative trajectories of
$\epsilon(t)$. For example, we consider quenches where $1/\epsilon \leq 1$ initially, so that the system
is in the normal state, and then
$1/\epsilon(t)$ is smoothly tuned to a value
arbitrarily close to the critical point, but still in the disordered phase.

We also consider quenches where $1/\epsilon(t)$ is tuned from a small value $\leq 1$, to a value inside the ordered phase, where it becomes
negative for a certain period of time,
before being tuned back again to its initial value. Thus for this quench, the system parameters are returned back to that of a normal
electron gas.
We present results for how the optical conductivity evolves in time for these two quench protocols. Our approach can be generalized
to other protocols.

We find strong signals in the optical conductivity driven by the interplay of critical slowing down
and fluctuation conductivity effects. In particular the fluctuation effects must be understood in an inherently non-equilibrium way.
This last point may be seen by comparing time-scales. A na\"{i}ve scale for fast interacting processes is
$\hbar/T \sim 10^{-2} \text{ps}$  at room temperature. Thus, if a transient state exists for  picoseconds~\cite{Mitrano15},
many interactions will occur over the life of the state, perhaps implying some sort of effective equilibrium, and therefore
a justification of using a temperature $T$ to describe the electron gas.

However, there is a much slower scale near criticality: the lifetime of superconducting fluctuations, given by $\sim \hbar/T \epsilon$.
For reasonable trajectories $\epsilon(t)$, we
encounter situations where the fluctuations relax slower than
$\epsilon(t)$ changes, and therefore the dynamics cannot be described by an adiabatic approximation.
As a result, one may have
$T < T_c(U(t))$ for a period of time $t$, yet true long range order may not have developed.

One may even provide a lower bound on the time $t$ for which  $T<T_c(U(t))$
instantaneously (or $\epsilon(t)<0$), yet true long range order does not develop.
We estimate $t$ as the time at which superconducting fluctuations grow to such an extent that they begin
to interact strongly, a necessary intermediate step before true long range order develops.
In equilibrium, the criterion for weakly interacting fluctuations is known as the
Ginzburg-Levanyuk criterion~\cite{Larkin00}. In this paper, we generalize
the Ginzburg-Levanyuk criterion to the case of a quench. From this we show that, for example for
a rapid quench into the ordered phase, the time for the violation of the Ginzburg-Levanyuk criterion is
$t\sim (T_c - T)^{-1}\log E_F \tau$, where $E_F$ is the Fermi energy. Thus for this period of time, even though the instantaneous $T_c$ is such that
the system can order in equilibrium, yet the system has not had time to develop long range order.

For a generous $T_c = 100$K, $T_c - T = 5$K, and $E_F\tau/\hbar =100$,
$t\sim 1 \text{ps}$. Thus a short lived experiment~\cite{Mitrano15} may never detect true superconductivity even though $T< T_c$.
Despite that, the superconducting fluctuations
in this transient regime can have important effects on the conductivity. Our goal is to identify these effects.

We briefly review
the equilibrium theory of fluctuation
superconductivity~\cite{Larkin00}. This theory describes how the conductivity of a dirty metal is corrected when $\epsilon \ll 1$.
In this region the electron-electron interactions are dominated by large superconducting fluctuations, even though there is no long range superconducting order.
This affects the conductivity through three channels. The first is the density of states effect, where incoherent scattering off of the superconducting
fluctuations renormalizes the single-electron propagation. In a dirty superconductor this is the weakest effect and negligible.
The second effect is the Azlamazov-Larkin (AL) effect~\cite{AL68-1,AL68-2} where the charged fluctuations serve as an additional channel
carrying current. This gives a positive contribution to the conductivity which, in equilibrium,
goes as $\epsilon^{-1}$ in dimension $d=2$.

The third effect is the Maki-Thompson (MT) effect~\cite{Maki68-1,Maki68-2,Thompson68}. Here an electron diffuses along a random trajectory until
it Andreev reflects off of a fluctuation. The resulting hole then diffuses through the same random trajectory in reverse.
As the electron and hole have opposite energy, the phase accumulated on the two random trajectories cancel. Thus all trajectories that
contribute to this process sum without destructive interference. The only limit on the number of contributing trajectories along which the
electron diffuses is inelastic collisions that destroy the phase coherence between the electron and hole trajectory.
The time for such processes, the coherence time $\tau_\phi$, is not controlled directly by the fluctuations and may be very long, $T\tau_\phi/\hbar \gg 1$.

We extend these results to the case of a non-equilibrium $\epsilon(t)$ caused by a time varying electron-electron interaction.
The conductivity correction cannot be obtained by taking the equilibrium calculation with time varying $\epsilon(t)$, for two reasons.
Firstly, as discussed, critical slowing down means that the size of the fluctuations at time $t$ is not given by $\epsilon(t)$. Secondly,
the AL and MT effects are long-lived processes, meaning that they occur on the scale of $1/T\epsilon$ and $\tau_{\phi}$ respectively.
As $\epsilon(t)$ is changing on these time scales, it is not even clear at what time $\epsilon(t)$ or the fluctuation density should be evaluated.
Finally we note that the actual experimentally measured quantity is a convolution of these various time-dependent quantities against the probe electric
field signal. Our calculation takes all of these into account.

Our results rest on several assumptions. (i) Self interaction of the fluctuations are neglected, this is valid as long as
the fluctuations are not extremely large.
(ii)  The electron occupation numbers relax to a thermal distribution
on a time-scale much shorter than the fluctuation lifetime. (iii) The time dependent perturbation, e.g. the driving laser,
may be modeled by a time varying electron-electron interaction.  (iv) This interaction is smoothly varying on the scale $\hbar/T$.

The paper is organized as follows. In Section~\ref{sec1} we present the model, outline the approximations, and discuss the
principle ingredients of the Feynman diagrams, namely the Cooperon and the fluctuation propagators. In this section we
also discuss the Ginzburg-Levanyuk criterion, and use it to present a lower bound for the time required to develop long range order.
In section~\ref{cond} we derive the expressions for the fluctuation conductivity, while in section~\ref{results}
we present results for the conductivity. This section also discusses the results where one rapidly quenches
to the critical point.  We show that an absence of any energy-scale in the problem results in
the optical conductivity showing power-law in time features, reminiscent of aging.
In section~\ref{conclu} we present our conclusions. Technical details are relegated to the appendices.

\section{Model} \label{sec1}

\begin{figure}
\includegraphics[width = \columnwidth]{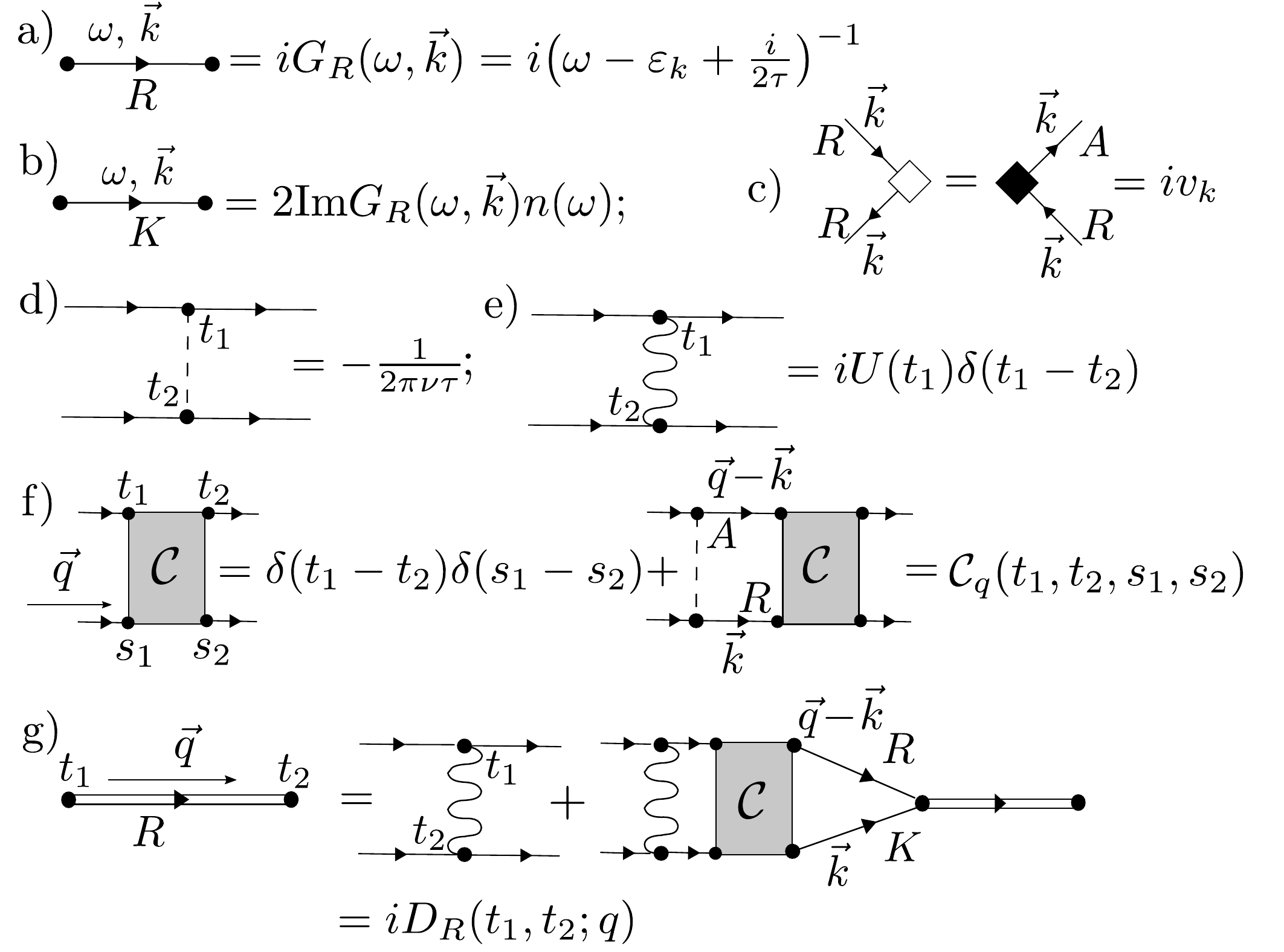}
\includegraphics[width = \columnwidth]{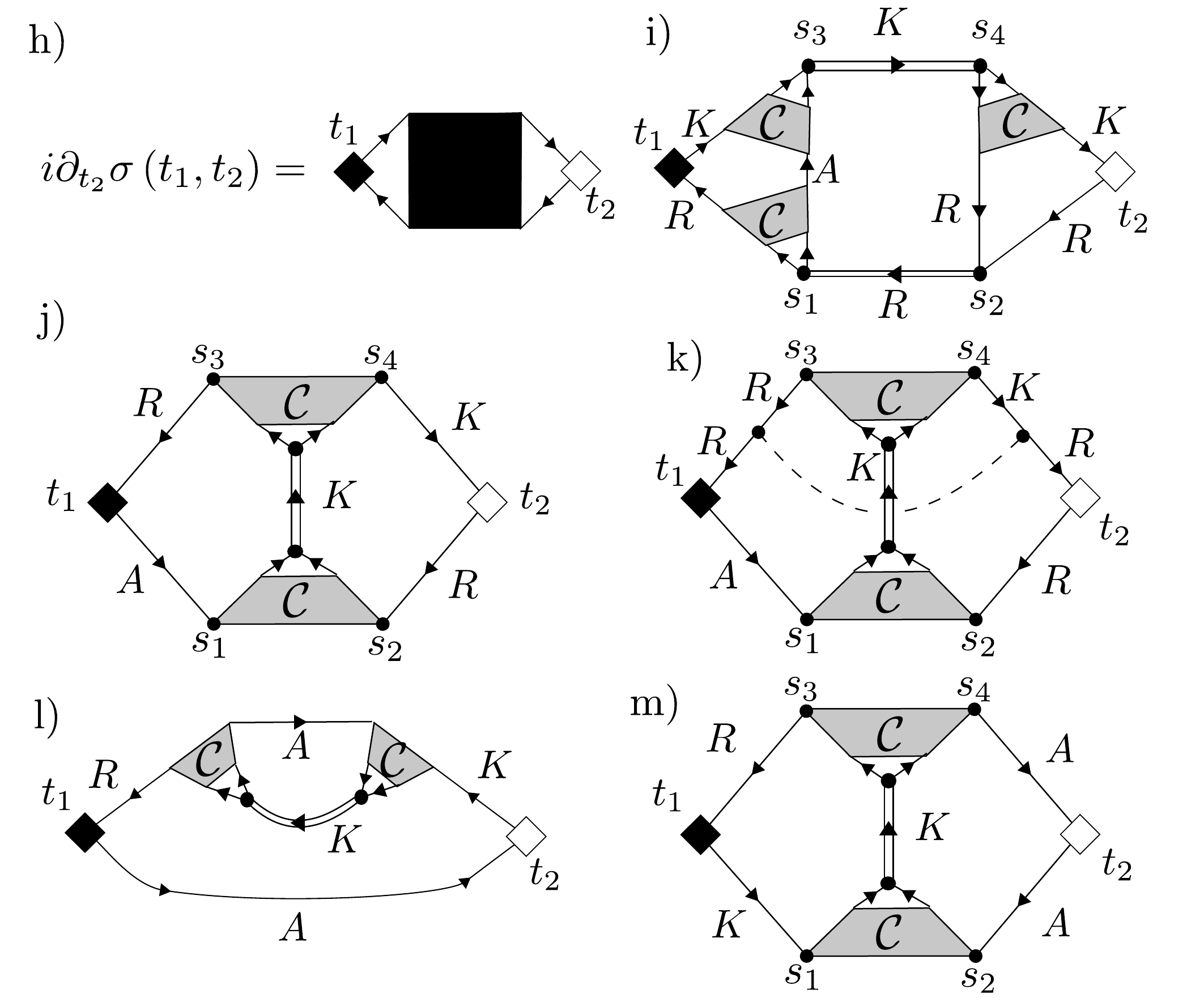}
\caption{Elements of the diagrammatic technique. a) Retarded electron Green's function $G_R$, including finite lifetime $\tau$ due to impurity scattering.
b) Keldysh electron Green's function $G_K$, with $n(\omega) = \tanh \omega/2T$. c) Current vertices. d) Impurity scattering vertex. e) Electron-electron interaction
vertex. The strength of the interaction is allowed to depend on time. f) Cooperon impurity ladder. g) Retarded fluctuation propagator $D_R$. The Keldysh
fluctuation propagator $D_K$, not pictured, is constructed analogously. h) Kubo formula expansion for the conductivity.
The black square represents the sum over all possible diagrams.
i) A diagram of AL type. j,k) Two diagrams of anomalous  MT type (these are identical in the Matsubara notation).
l,m) Two subleading diagrams: l) a density of states correction and m) a regular MT correction.
\label{fig:feyn}}
\end{figure}

The Hamiltonian is
\begin{eqnarray}
&&H  =  \sum_{kk's}\!\bigg[\! \left(\varepsilon_k\delta_{kk'}\! +\! V_{k-k'}\right)c^\dagger_{ks} c_{k's}\nonumber\\
&&+ U(t)\sum_{qs'}\! c^\dagger_{ks} c_{k-q s} c^\dagger_{k'-q s'}c_{k's'}\bigg].
\end{eqnarray}
Here $c_{ks}, c^\dagger_{ks}$ are the creation and annihilation operators of electrons with momentum $k$ and spin $s$,
$\nu$ is the density of states, and $U(t)$ is the time dependent interaction constant.
The potential $V$ is a Gaussian random potential with moment
\begin{equation}
\langle V_q V_{-q'}\rangle = \delta_{qq'}/2\pi\nu\tau.
\end{equation}
The conductivity is given by the Kubo formula
\begin{equation}
\sigma(t,t') = i\int_{-\infty}^{t'}ds\langle[J(t),J(s)]\rangle,
\end{equation}
where the current operator is
$J = \sum_k \frac{\partial \varepsilon_k}{\partial k} c^\dagger_k c_k$, and $\langle\cdot\rangle$ signifies both quantum and impurity averaging.
Since the system is not time translationally invariant $\sigma$ depends on both $t'$ and $t$ separately.
The calculation proceeds by evaluating the appropriate diagrams in the Keldysh technique summarized in Fig.~\ref{fig:feyn},
employing the standard diagrammatic technique reviewed in Ref.~\onlinecite{Kamenevbook2011}.

This calculation may appear very difficult given the underlying time dependence. However the problem
is greatly simplified by the existence of two sets of timescales.
The first are the "fast" time scales given by $\tau$, $T^{-1}$. These are much shorter than the second set of "slow" time scales given by
$\hbar/ T\epsilon(t) = \hbar/[T-T_c(t)]$ and $\tau_\phi$.
Since we are interested in dynamics controlled by long time scales, functions $f(t,t')$ which are supported only on short time scales may be \emph{e.g.} expanded
in derivatives $f(t,t') \approx f_0(t)\delta(t-t') + f_1(t)\partial_t\delta(t-t')$ etc. We explain this further below, and in
the appendices, where we outline the calculations in detail.

\subsection{Electron Green's function, Cooperons, and Fluctuation propagators}
In this section we discuss the ingredients of the calculations, namely the electron Green's functions $G_{K,R}$, the Cooperon $\mathbb{C}$, and
the superconducting fluctuation propagators $D_K,D_R$. These are all shown diagramatically in Fig.~\ref{fig:feyn}.

First let us discuss the electron Green's functions $G_K,G_R$.
In the regime of a good metal $E_F \tau \gg 1$, we may neglect all diagrams with crossing impurity lines in the Dyson
equation, giving the usual metallic Green's function,
\begin{align}
G_R(\omega, k) &= \left(\omega -\varepsilon_k + \frac{i}{2\tau}\right)^{-1},
\label{eq:defGR} \\
G_K(\omega, k) &= \left[G_R(\omega,k) - G_A(\omega,k)\right] \tanh\frac{\omega}{2T},
\label{eq:defGK}
\end{align}
where we are taking the electron distribution function $G_K$ to be it's equilibrium value at fixed temperature. The rationale is that
the fermion distribution thermalizes at times of the order of $\hbar/T$. On the other hand, the quantity that we are interested in, namely the conductivity,
is strongly affected by the superconducting fluctuations. As we show below, these
thermalize on much
longer time-scales. We note in passing that including heating effects via a time-varying temperature is straightforward,
as $G_K(t,t')$ is only supported when $|t'-t| \sim T^{-1}$.

The other significant diagram is the Cooperon $\mathbb{C}_q(t_1,t_2;s_1,s_2)$, given in Fig.~\ref{fig:feyn}(f).
This represents the previously mentioned set of processes where in the time  $t_1$ to $t_2$ an electron diffuses along a random trajectory, and a
hole diffuses on the same trajectory in reverse, in the time $s_2$ to $s_1$. In the language of Green's functions, the Cooperon therefore
connects $G_R$ and $G_A$ Green's functions, but never two $G_R$ or two $G_A$.

The Cooperon obeys the equation
\begin{align}
\mathbb{C}_q(t_1,t_2;s_1,s_2) &= \delta(t_1-t_2)\delta(s_1-s_2)\nonumber\\
 &\hspace{-.5in}+ \int dt'ds' \mathcal{P}_q(t_1,t';s_1,s') \mathbb{C}_q(t',t_2;s',s_2),
\label{eq:defC}
 \\
\mathcal{P}_q(t,t';s,s') &= \frac{1}{2\pi\nu\tau}\sum_k G_R(t,t';-k + q)G_A(s,s';k ).
\label{eq:defP}
\end{align}

As shown in Appendix~\ref{Cder}, as long as we are exploring times such that $t-t',s'-s\gg T,\tau^{-1}$ in Eq.~\eqref{eq:defP},
we may solve for $\mathbb{C}_q$ by Fourier transforming Eq.~\eqref{eq:defC} obtaining,
\begin{align}
\mathbb{C}_q(\omega_1,\omega_2)\simeq \frac{1/\tau}{Dq^2 + \frac{1}{\tau_{\phi}}-i(\omega_1-\omega_2)}.\label{Cwdef1}
\end{align}
The above implies (see Appendix~\ref{Cder}) that in time, the Cooperon has the form,
\begin{align}
\mathbb{C}_q(t_1,t_2,s_1,s_2) &\approx \theta\!\left(t_1-t_2\right)
\frac{1}{\tau}e^{-\frac{1}{2}\left(t_1-t_2+s_2-s_1\right)\left(\frac{1}{\tau_\phi} + D q^2\right)}\nonumber\\
	&\times \delta\!\left(t_1  - t_2 + s_1 - s_2\right).
\label{eq:longtimeC}
\end{align}
Above $\tau_{\phi}$ is the phenomenological phase breaking time. In its absence the low frequency, long-wavelength limit of
the Cooperon diverges. The reason for this divergence was explained in the Introduction. It is due to a proliferation in the number
of diagrams where the electron and the Andreev reflected hole diffuse phase coherently along identical trajectories.

We now consider the superconducting fluctuation propagator.
It is sufficient to consider only the ladder diagrams, Fig.~\ref{fig:feyn} (g). This neglects any of
the self-interaction of the fluctuations, and is justified as long the fluctuations are not "too large" -
the precise criterion is given below.

We consider the object $\Pi_R$ multiplying $D_R$ on the right-hand side of Fig.~\ref{fig:feyn}(g),
\begin{align}
\Pi_R(u_1,u_2;q) \equiv \sum_k\int\!\! dt' ds'\,\, \mathbb{C}_q\!
\left(u_1,t';u_1,s'\right)
\nonumber\\
\times G_R(t',u_2;k)G_K(s',u_2;q-k).
\end{align}
This is exponentially suppressed unless $|s'-u_2|,|t'-u_2| \lesssim \tau,T^{-1}$.
As the system is time translation invariant on this scale,
we Fourier transform
\begin{align}
\Pi_R(\omega, q) &= i\int \frac{d\omega_1}{2\pi}\mathbb{C}_q(\omega_1,-\omega_1+\omega) \nonumber\\
&\times \sum_k G_R(\omega_1,k) G_K(-\omega_1+\omega,-k+q),
\end{align}
and obtain the standard result~\cite{Larkin00},
\begin{align}
\Pi_R(\omega, q)/\nu = \chi(T) + i\frac{\pi \omega}{8 T} + \xi^2 q^2,
\label{eq:defPiR}
\end{align}
where the coherence length
\begin{align}
\xi = \sqrt{\pi D/ 8 T},
\end{align}
in the dirty limit, $D$ is the Diffusion constant, and $\chi(T)$ is a constant $\propto \ln(T/E_F)$.
Now using that the Dyson equation, $D_R$ within the ladder approximation is
\begin{align}
D_R^{-1}(t,t')= \delta_{t,t'}U^{-1}(t)-\Pi_R(t,t'),
\end{align}
leading to the following equation of motion for the retarded  propagator $D_R$ in time space,
\begin{gather}
\left[\frac{\pi}{8 T}\frac{\partial}{\partial t} + \epsilon(t) + \xi^2 q^2\right] D_R(q,t,t') = \frac{1}{\nu}\delta(t-t'),
\label{eq:defDR}
\end{gather}
where the time dependent detuning is related to the interaction by
\begin{align}
\epsilon(t) \approx \frac{1}{\nu U(t)} - \chi(T).
\end{align}

Similarly one may derive that the Keldysh part of the fluctuation propagator $D_K$.
This quantity is symmetric in the time-indices, $D_K(t,t') = D_K(t',t)$, and obeys the equation of
motion,
\begin{align}
D_K = D_R\circ\Pi_K\circ D_A,
\end{align}
where $\circ$ denotes convolution.
On the time-scales of interest, $\Pi_K$ takes its thermal equilibrium value $\Pi_K(t,t')= \left(\pi\nu/2\right) \delta(t-t')$.
This leads to an equation of motion,
\begin{equation}
iD_K(t,t') = 4\biggl[D_R(t,t') B_q(t') + B_q(t)D_A(t,t')\biggr],\label{Bqdef}
\end{equation}
where we have introduced the dimensionless fluctuation density $B_q$. Employing the equation of motion for $D_R$ derived above,
and the definition Eq.~\eqref{Bqdef}, $B_q$ is found to obey the differential equation:
\begin{equation}
\left[\frac{\partial}{\partial t} + \frac{16T}{\pi}\left(\epsilon(t) + \xi^2 q^2\right)\right] B_q(t) = T.
\label{eq:kineticB}
\end{equation}

The physical meaning of $B_q$ is that it is the occupation probability of the superconducting fluctuations.
The above equation, for a time-independent $\epsilon$, leads to the thermal equilibrium value for $B_q$,
\begin{align}
B_q^{\rm eq}= \frac{\pi/16}{\epsilon +\xi^2 q^2},
\end{align}
with $D_K, D_R$ obeying the classical Fluctuation dissipation theorem,
\begin{align}
&D_K(q,\omega) = \frac{2T}{\omega}\biggl[D_R(q,\omega)-D_A(q,\omega)\biggr], \\
&D_R(q,\omega) = \frac{-1/\nu}{i\frac{\pi \omega}{8 T}-\left(\epsilon + \xi^2 q^2\right)}.
\end{align}
Recall that $B_q$ has the characteristic $1/q^2$ behavior at the critical point $\epsilon=0$.

\subsection{Ginzburg-Levanyuk criterion generalized to quench dynamics}

We now discuss the selection of the diagrams and the region of applicability. In equilibrium one may derive the appropriate condition by comparing $e.g.$ the fluctuation
correction to the specific heat with the jump in the specific heat at the critical point~\cite{Larkin00}. Alternatively one may compare the DC
correction to the conductivity with the Drude form. Both these lead to the Ginzburg-Levanyuk criterion $\epsilon \gg 1/E_F\tau$.
This is not satisfactory in our strongly non-equilibrium setting where $\epsilon(t)$ is not directly connected to the size of the fluctuations, and where there is no single notion of a DC conductivity. Instead we must express the criterion directly in terms of the fluctuation strength.

We carry out this analysis in Appendix~\ref{glc}. There we explicitly show that the selection of
diagrams is governed by the smallness of $B_q(t) \ll E_F\tau$. This reduces to the usual
Ginzburg-Levanyuk~\cite{Levanyuk59} criterion in equilibrium. Now let us briefly discuss the consequences of this condition
$B_q(t) \ll E_F\tau$.

Consider a quench $\epsilon(t)  = -\epsilon_f < 0$ for $t>0$ and $\epsilon \sim 1$ for times $t<0$. The solution of
Eq.~\eqref{eq:kineticB} gives that,
\begin{equation}
B_q\left(t\right) = \frac{\pi}{16}\frac{1-e^{- 16T(q^2\xi^2 -\epsilon_f)t/\pi}}{q^2\xi^2-\epsilon_f}.\label{Bq1}
\end{equation}
Note that $B_q(t=0)=0$ consistent with a large initial detuning $\epsilon \sim 1$, and therefore vanishingly small superconducting
fluctuations in the initial state. With time $t$, the fluctuations grow, reaching their thermal equilibrium value, consistent with the
the final detuning. However, if $\epsilon(t) < 0$ as in the above example, the fluctuations may grow to an extent such that our starting approximation
of neglecting self-interactions between the fluctuations may breakdown. We now estimate the time for this to happen.

Since the $q=0$ mode grows the fastest, the violation of the Ginzburg-Levanyuk criterion occurs at a time $B_{q=0}(t)=E_F\tau$. Substituting
Eq.~\eqref{Bq1} into this condition gives, to logarithmic accuracy,
\begin{align}
Tt \sim \frac{1}{\epsilon_f}\ln\biggl(E_F\tau\biggr),
\end{align}
where $\epsilon_f\equiv (T_c-T)/T$. This is precisely the condition presented in the Introduction.

\begin{figure}
\includegraphics[width = \columnwidth]{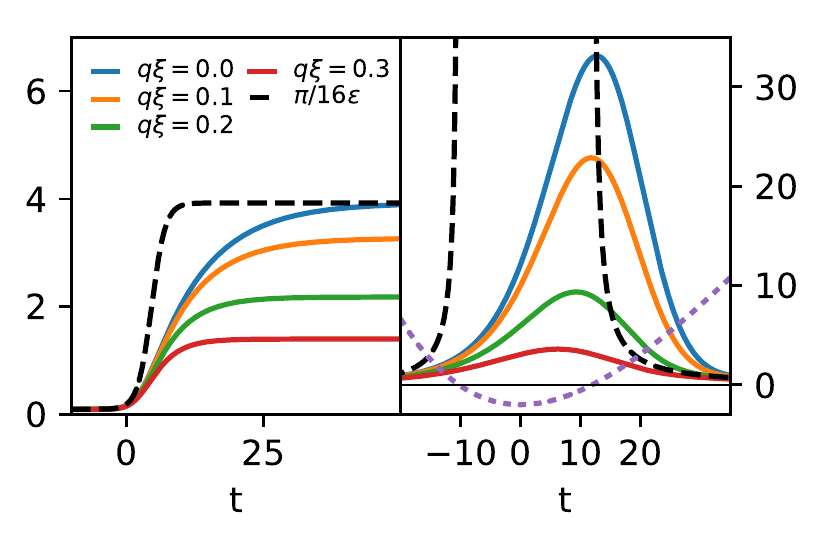}
\caption{The growth of the dimensionless superconducting fluctuation density $B_q$ (Eq.~\eqref{eq:kineticB})
at different lengths ($q^{-1}$ in units of the coherence length $\xi$), and under a changing detuning $\epsilon(t)$ from the superconducting critical point.
The time $t$ is given in units of $\pi\hbar/8T$. The dashed line shows $\pi/16\epsilon(t)$ which equals $B_{q=0}$ in equilibrium.
The dotted purple line in the right panel gives $\epsilon(t)$ at arbitrary scale. Left and right panels show two different quench protocols.
In the left panel the detuning saturates at the value
$\epsilon = 0.05$. In the right panel, the detuning is $\epsilon(t) = \epsilon_0 + (\epsilon_{\rm min} - \epsilon_0)(t/t_* e)\exp(-t/t_*)\theta(t)$
with the parameters $t_* = 30$, $\epsilon_0 = 1$, $\epsilon_{\rm min} = -0.05$.
\label{fig:flucs}
}
\end{figure}

The solution of $B_q(t)$ for two different quench profiles are shown in Fig.~\ref{fig:flucs}. The system is always initially in the normal
phase ($\epsilon \geq 1$).
The left panel shows the case of a
quench where the detuning is smoothly tuned to a final value close to the critical point, and kept fixed there.
The right panel on the other hand shows the case where the detuning is tuned to a negative value for a certain amount of time, and
eventually returned
back to the starting positive value. Thus in this trajectory, the system traverses what would
have been the ordered phase in equilibrium, for a certain duration of time.

The plots for $B_q$ are shown for a three different momenta $q$. Note that $B_{q=0}$ in thermal equilibrium equals $\pi/16\epsilon$.
Since the system thermalizes, $B_{q=0}$ coincides with $\pi/16\epsilon(t)$ at long times, in the figure.

It is interesting to note if interactions are added in Eq.~\eqref{eq:kineticB} via a cubic term in $B_q$, this equation becomes identical to that studied for classical quenches of the bosonic O(N) model~\cite{Gambassi2005} and quantum quenches of the dissipative O(N) model~\cite{Gagel2014}. On the other hand,
quantum quenches of the closed bosonic O(N) model
correspond to changing $\partial_t \rightarrow \partial_t^2$, with the self-interactions accounted for
within Hartree-Fock~\cite{Sotiriadis2010,Sondhi2013,Smacchia2014,Maraga2015} and renormalization group methods~\cite{Chiocchetta2015,Tavora16}.

In the problem studied here, even though the dynamics of $B_q$ is essentially free, it affects the
fermions to which $B_q$ is coupled in non-trivial ways. We discuss this effect in the next section.

\section{Conductivity} \label{cond}
With this description of the dynamics of the fluctuations, we now calculate the corrections to $\sigma(t,t')$ caused by them.
This is the sum of diagrams indicated schematically by the sum of all diagrams of the form Fig.~\ref{fig:feyn}(h).
Some representative diagrams are shown in Fig.~\ref{fig:feyn}(i-m).

The diagrams are chosen analogously to the equilibrium calculation. First, as we are interested in the effect of large fluctuations,
we include only diagrams containing $D_K$. As $D_K \sim B_q$ grows with  small $\epsilon(t)$, this is sufficient to produce a singular correction.
However this effect alone is somewhat weak. For example, in equilibrium $B_q \propto (\xi^2 q^2 +\epsilon)^{-1}$ and integrating over $d^2 q$ gives only
$\log \epsilon$.

Certain diagrams have an additional enhancement. This comes from the fact that the Cooperon and fluctuation propagator is long lived,
so that the correction to the conductivity does not vanish
for $|t_1-t_2| \gg T^{-1},\tau$ but rather is supported up to the long time scales $\tau_\phi$, $1/T\epsilon(t)$.
This may be seen in Fig.~\ref{fig:feyn} (i) where $t_1$ and $t_2$ are connected only by long lived fluctuation lines.
Similarly, in the anomalous MT contribution Fig.~\ref{fig:feyn} (j,k), the slow decay of the Cooperon with time means that $t_1-t_2$ may be large.
Thus when the low-frequency conductivity is calculated, there is an additional enhancement from integrating over $|t_1-t_2|\gg \tau,T^{-1}$.
The consequence of this in equilibrium is that these diagrams diverge as $1/\epsilon$ as $\epsilon \rightarrow 0$.

In the density of states contributions, Fig.~\ref{fig:feyn} (l), and the regular MT corrections, Fig.~\ref{fig:feyn} (m), this long-livedness is absent
and hence they are negligible compared to the larger AL and anomalous MT contribution. This can be easily seen for diagram Fig.~\ref{fig:feyn} (l)
where the electron Green's function $G_A(t_2,t_1)$ restricts the contributions to short times $|t_1-t_2|\leq \tau$. How the regular MT
diagram in Fig.~\ref{fig:feyn} (m) differs from the anomalous MT in Fig.~\ref{fig:feyn} (j) is discussed in Appendix~\ref{kuboMT}.
At intermediate impurity  concentration  $T\tau \sim 1$, all these diagrams become comparable to the MT correction~\cite{Varlamov00}. We do not
discuss this case here.

Therefore the contributing diagrams are those of the form Fig.~\ref{fig:feyn} (i,j,k). These are equivalent to the most divergent equilibrium diagrams,
excepting for some superficial differences owing to the Keldysh technique, \emph{e.g} diagrams (k) and (j) are represented in one diagram in the
Matsubara technique. Moreover, diagram Fig.~\ref{fig:feyn}(i) is one of several diagrams. Since the impurity line cannot connect the two retarded propagators in the right fermion loop, the diagram shown has only 3 Cooperons.
Some of the others that are not shown do have 4 Cooperons. All possible diagrams are discussed in the Appendix~\ref{kuboAL}.

Evaluating the diagrams, we obtain for the AL and anomalous MT diagrams,
\begin{subequations}
\begin{align}
\sigma^{\rm AL}(t_1,t_2) &= 32\int_{-\infty}^{t_2}\!\!\!\! ds\int\!\! \frac{d^2 q}{(2\pi)^2}\, \xi^4 q^2 \nu^2 |D_R(q,t_1,s)|^2 B_q(s),
\label{eq:sigAL}\\
\sigma^{\rm MT}(t_1,t_2) &= 8 D \!\! \int\!\!\! \frac{d^2q}{(2\pi)^2} B_q\left(\frac{t_1+t_2}{2}\right)
 e^{-\left(\frac{1}{\tau_\phi} + D q^2\right)\left( t_1-t_2\right)}.
 \label{eq:sigMT}
\end{align}
\end{subequations}
Note that as in equilibrium, the AL contribution in Eq.~\eqref{eq:sigAL} has two fluctuation propagators contributing to the conductivity, these
are $D_KD_R$ which using our parameterization in Eq.~\eqref{Bqdef} we write as $\equiv |D_R|^2B_q$. Further technical details of the Keldysh calculation
for $\sigma^{\rm AL}$ and $\sigma^{\rm MT}$ are presented in Appendices~\ref{kuboAL},~\ref{kuboMT}.
Moreover, in Appendix~\ref{condeq}, we show that these expressions reduce to the known results in equilibrium when the detuning $\epsilon$
is time-independent.

We note that $\sigma^{\rm MT}(t_1,t_2)$ has the peculiar property that the fluctuations are evaluated not at $t_1$ or $t_2$ but the average time $(t_1 +t_2)/2$.
Physically, the electron must be excited, diffuse, Andreev reflect, diffuse back and be absorbed. As the two diffusions must be identical,
the fluctuations are interacted with precisely halfway in time between the excitation and measurement.

\section{Results} \label{results}
We now consider the appearance of $\sigma=\sigma^{\rm AL}+\sigma^{\rm MT}$ for certain choices of $\epsilon(t)$.
A particularly interesting choice is the critical quench, where $\epsilon(t)$
switches instantaneously from $\epsilon(t) = \epsilon_i \geq 1 $ for $t <0$ to $\epsilon(t) = 0 $ for $t>0$.
Substituting for the fluctuation propagators for the critical quench, we have,
\begin{align}
\sigma(t_1,t_2) &= \frac{2T}{\pi} \bigg[-\frac{t_2}{t_1}- \log\left(1- \frac{t_2}{t_1}\right)\nonumber\\
&\,+  \frac{1}{2}e^{-(t_1-t_2)/\tau_\phi} \log\left(1+ \frac{t_1+t_2}{t_1-t_2}\right)\bigg].
\end{align}
The above result is valid when $T(t_1-t_2) \gg 1$.  When either $t_1-t_2\gg \tau_\phi$ or $t_1-t_2 \ll \tau_\phi$ this becomes solely a function of $t_2/t_1$. Such a power law dependence on $t_2/t_1$, is a classic example of aging.
This power-law scaling behavior is a consequence of the "criticality" of the quench when $\epsilon = 0$. In this case, the only "slow" scale is $\tau_\phi$.
While previous studies of bosonic O(N) models showed aging after a critical quench~\cite{Maraga2015,Chiocchetta2015,Tavora16},
our results are an example of aging for critical quenches involving fermions.

\begin{figure}
\includegraphics[width = 3.25in]{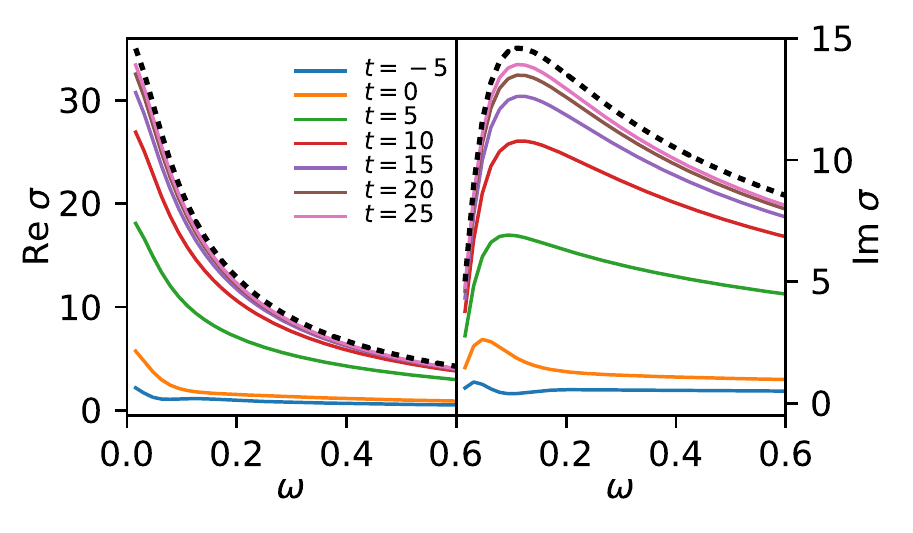}
\caption{Conductivity $[e^2/\hbar]$ as a function of frequency $\omega$ for several times $t$. The left panel shows real part, the right panel shows
the imaginary part.
$t, \omega^{-1}$ are in units of $\pi\hbar/8T$. The detuning $\epsilon$ varies according to Fig.~\ref{fig:flucs},
left panel and $\tau_\phi = 20\times(\hbar\pi/8T)$. The dashed line gives the equilibrium result for the final value of the detuning
$\epsilon = .05$.
\label{fig:saturation}
}
\end{figure}

\begin{figure}
\includegraphics[width = \columnwidth]{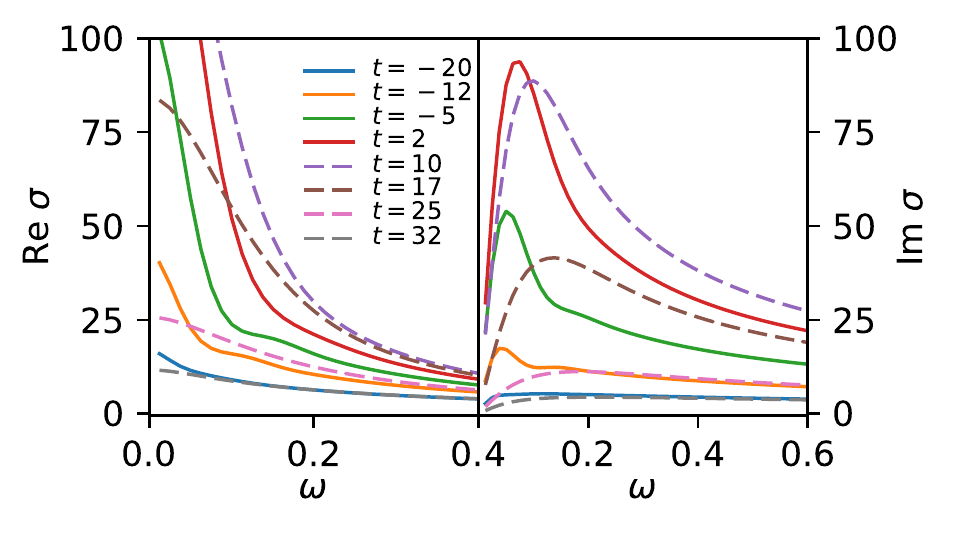}
\caption{Conductivity $[e^2/\hbar]$ as a function of frequency $\omega$ for several times $t$. $t,\omega^{-1}$ are
in units of $\pi\hbar/8T$. The detuning is given in
Fig.~\ref{fig:flucs}, right panel and $\tau_\phi = 20 \times(\hbar\pi/8T)$.  In order to improve clarity, lines for $t\leq 2$ are shown with
full lines and those for $t>2$ are shown with dashed lines. The conductivity $\sigma$ as  $\omega\rightarrow 0$ reaches a maximum of $\sim 175$ at $t\approx 7$.
\label{fig:smoothquench}
}
\end{figure}

We now consider how the change in conductivity $\sigma = \sigma^{\rm AL} +\sigma^{\rm MT}$ would appear in measurement. For simplicity we plot a time dependent
optical conductivity given by $\sigma(\omega,t) = \int\!d\tau\,e^{i\omega \tau} \sigma\!\left(t+\tau,t\right)$. This is shown for the two quenches in
Figs.~\ref{fig:saturation} and \ref{fig:smoothquench}.

Fig.~\ref{fig:saturation} displays $\sigma(\omega,t_0)$ for the quench profile of Fig.~\ref{fig:flucs}, left panel. For this case
$\epsilon$ is tuned smoothly to a value close to the critical point, but still on the disordered side, and kept fixed at that final value.
The long time equilibrium result (dashed) has the $\text{Re}\sigma(\omega)$ increasing as $\omega\rightarrow 0$ until the slowest frequency
$\min(\tau_\phi^{-1}, \epsilon T)/\hbar$ is reached. Below this frequency, the curve flattens to its DC value.
Correspondingly there is a peak in $\text{Im}\sigma$ at this same slowest frequency. In the time dependent results
(full lines) we may directly see critical slowing down and thermalization.
In particular, the conductivity converges to its equilibrium value but only on a slow time scale $\approx 20\times  (8T/\pi)$,
significantly wider than the underlying quench.

Fig.~\ref{fig:smoothquench} shows $\sigma(\omega,t_0)$ for a quench profile shown in the right panel of Fig.~\ref{fig:flucs}. Here the trajectory
involves $\epsilon$ smoothly entering the ordered phase and leaving it.
This trajectory is perhaps
more appropriate to a solid state system, where a pump laser smoothly
changes the interaction with time, first increasing it and then decreasing it back to its initial value. In particular we give the case where instantaneously
$\epsilon(t) < 0$ but the size of fluctuation may remain within
the $B_q(t)\ll E_F\tau$ limit.

Although the conductivity increases and decreases as expected, we make the following observations. (i) the peak of the conductivity lags the minimum of the detuning,
(ii) the actual dependence of $\sigma(\omega)$ on $\omega$ is not given by any equilibrium choice of parameters. Indeed the profiles of the conductivity
on the ramp-up are markedly different than on the ramp down. This is not surprising as the corrections are determined by the fluctuations,
and the fluctuations are determined by Eq.~\eqref{eq:kineticB} - and not any equilibrium distribution. 
The peak of the low frequency
$\text{Re}\sigma(\omega\rightarrow 0)$ is $\sigma \sim 175 e^2/\hbar$.
This agrees in order of magnitude with the following estimate: setting an effective detuning by the relationship $\epsilon_{\rm eff} \equiv \pi/16 B_{q=0}$,
and using the equilibrium formula in Appendix~\ref{condeq} with $\epsilon_{\rm eff}$.
For the present quench with $\max B\approx 35$, $\epsilon_{\rm eff} \approx\times 10^{-3}$,
and $\text{Re}\sigma(\omega\rightarrow 0)\approx  60 e^2/\hbar$. This estimate shows that it is indeed the size of the fluctuations
which is controlling, at least qualitatively, the conductivity, and not the instantaneous value of $\epsilon(t)$.

The conductivity $\sigma(t,t')$ discussed above, is defined by the relationship $J(t) = \int dt'\sigma(t,t')E(t')$. However constructing
$\sigma(t,t')$ experimentally for arbitrary $t,t'$ is difficult.
If delta function pulses
$E\propto \delta(t-t_0)$ could be produced, then $J(t)$ could be measured for many such pulses and
$\sigma(t,t')$ directly reconstructed. However delta function pulses are not experimentally available.
Instead an electric field pulse $E(t) = f(t-t_0)$, centered at $t_0$, is applied, and the induced current $J(t)$ is
(indirectly) measured. Following this, $J$, $E$ are Fourier transformed and
an experimentally motivated conductivity~\cite{Kennesprb17} "$\sigma(\omega,t_0)$" $= J(\omega)/E(\omega)$ is extracted.
In Appendix~\ref{sigE}, this experimental definition
of $\sigma(\omega,t)$ is considered.

\section{Conclusions} \label{conclu}
In summary, we make predictions for the time-resolved optical conductivity due to non-equilibrium superconducting fluctuations. Although
the system never develops long range order, the fluctuations provide an additional low resistance current carrying channel, which
causes the imaginary part of the low frequency optical conductivity to get transiently enhanced.
This is also what is seen in experiments~\cite{Mitrano15}.

Note that if the system was simply becoming a good conductor transiently, the imaginary part would behave in a similar way. However
a detailed theory like the one presented here, with explicit dependence of the real and imaginary parts of
 the conductivity on detuning, $\tau_{\phi}$, and temperature, will
provide a guide for future experiments to quantitatively confirm the superconducting nature of the
transient state.

Our calculations also suggest that experiments can be used to extract information on the underlying trajectory
of $\epsilon(t)$, and hence of the light-induced interaction. This would be a more direct study of the critical superconducting dynamics,
rather than indirectly via \emph{e.g.} Kibble-Zurek effects~\cite{KZrev14}. It should also be emphasized that the MT and AL effects,
which have similar footprints in DC measurements ($\sim 1/\epsilon$, see Appendix~\ref{condeq}), behave fundamentally differently in a time-resolved situation.
As the two contributions are differently sensitive to \emph{e.g.} pairing symmetry, time-resolved fluctuation measurements might be
capable of elucidating the underlying superconducting order.

{\sl Acknowledgements:}
This work was supported by the US National Science Foundation Grant NSF-DMR 1607059.

\begin{appendix}

\section{Derivation of Eq.~\eqref{eq:longtimeC}} \label{Cder}

Employing the form of $G_{R,K}$ in Eq.~\eqref{eq:defGR},~\eqref{eq:defGK}, we may Fourier transform
$\mathcal{P}_q(t,t';s,s')$ in Eq.~\eqref{eq:defP} with respect to the times $t-t'$ and $s-s'$,
\begin{align}
\mathcal{P}_q(\omega_1,\omega_2) = \frac{1}{2\pi\nu\tau}\sum_k G_R(\omega_1,k + q/2)G_A(\omega_2,-k+q/2).
\end{align}
The above may be written as,
\begin{align}
\mathcal{P}_q(\omega_1,\omega_2) &= \frac{1}{2\pi\tau}\int d\xi
\biggl\langle \biggl(\frac{1}{\omega_1-\xi-\frac{\vec{v}_k\cdot \vec{q}}{2}
+\frac{i}{2\tau}}\biggr)\nonumber\\
&\times \biggl(\frac{1}{\omega_2-\xi+\frac{\vec{v}_k\cdot \vec{q}}{2}
-\frac{i}{2\tau}}\biggr)\biggr\rangle_{\hat{k}}
\end{align}
where $\langle \rangle_{\hat{k}}$ denotes the angular integral.
Performing the $\xi$ integral and noting that the integrand being peaked at $\xi=0$,
the angular integral is over the Fermi surface,
\begin{align}
\mathcal{P}_q(\omega_1,\omega_2) &= \biggl\langle \frac{1}{1 - i\tau(\omega_1-\omega_2)+i\tau\vec{v}_F\cdot \vec{q}} \biggr\rangle_{\rm FS}.
\end{align}

Expanding in $q$, and assuming that $\omega_1-\omega_2\ll 1/\tau$, we obtain,
\begin{align}
\mathcal{P}_q(\omega_1,\omega_2) &\simeq 1 + i\tau(\omega_1-\omega_2) -\tau^2\langle(\vec{v}_F\cdot \vec{q})^2\rangle_{\rm FS}\nonumber\\
&= 1 + i\tau(\omega_1-\omega_2) - Dq^2\tau,\label{Pq1}
\end{align}
where we  have used that $D = v_F^2\tau/2$ is the Diffusion constant in spatial dimension $d=2$.

For the Cooperon, as long as we are exploring time-scales long compared to $\tau,T^{-1}$,
one may solve for the Cooperon by Fourier transforming Eq.~\eqref{eq:defC}. Thus we obtain,
\begin{align}
\mathbb{C}_q(\omega_1,\omega_2)=\frac{1}{1-{\mathcal P}_q(\omega_1,\omega_2)}.
\end{align}
Using Eq.~\eqref{Pq1}, we obtain,
\begin{align}
\mathbb{C}_q(\omega_1,\omega_2)\simeq \frac{1/\tau}{Dq^2 + \frac{1}{\tau_{\phi}}-i(\omega_1-\omega_2)}.\label{Cwdef}
\end{align}
We have also introduced a phenomenological dephasing time $\tau_{\phi}$.

Fourier transforming back into time space,
\begin{align}
\mathbb{C}_q(t,s)&= \int \frac{d\omega_1}{2\pi}\int \frac{d\omega_2}{2\pi} e^{-i\omega_1t-i\omega_2 s}\mathbb{C}_q(\omega_1,\omega_2)\nonumber\\
&=\frac{1}{\tau}\theta(t-s)\delta(t +s) e^{-\frac{1}{2}(t-s)(D q^2 +\frac{1}{\tau_{\phi}})}.
\end{align}
This is Eq.~\eqref{eq:longtimeC} in the main text with $t=t_1-t_2,s=s_1-s_2$.

\section{Nonequilibrium Ginzburg-Levanyuk condition} \label{glc}

We discuss the selection of the diagrams and the region of applicability.
To do this we note that every diagram in the notation of Fig.~\ref{fig:feyn} may be decomposed in two parts: Fluctuation propagators and closed fermion loops,
(possibly containing current vertex insertions). We have seen that fluctuation propagators are of size at most $D_K(q) \sim B_q T/\nu$ and $D_R \sim T/\nu$. (Going forward we
suppress the $q$ subscript as it is expected that $B_q$ is largest at $q=0$).
Since $B\gg1$ we would like to have all propagators be of Keldysh type. However, this is not allowed by causality and we must have
at least one retarded
fluctuation line per fermion loop. Thus we may imagine that all lines are of Keldysh type, and then divide by $B$ for each fermion loop.

\begin{figure}
\includegraphics[width =3.4in]{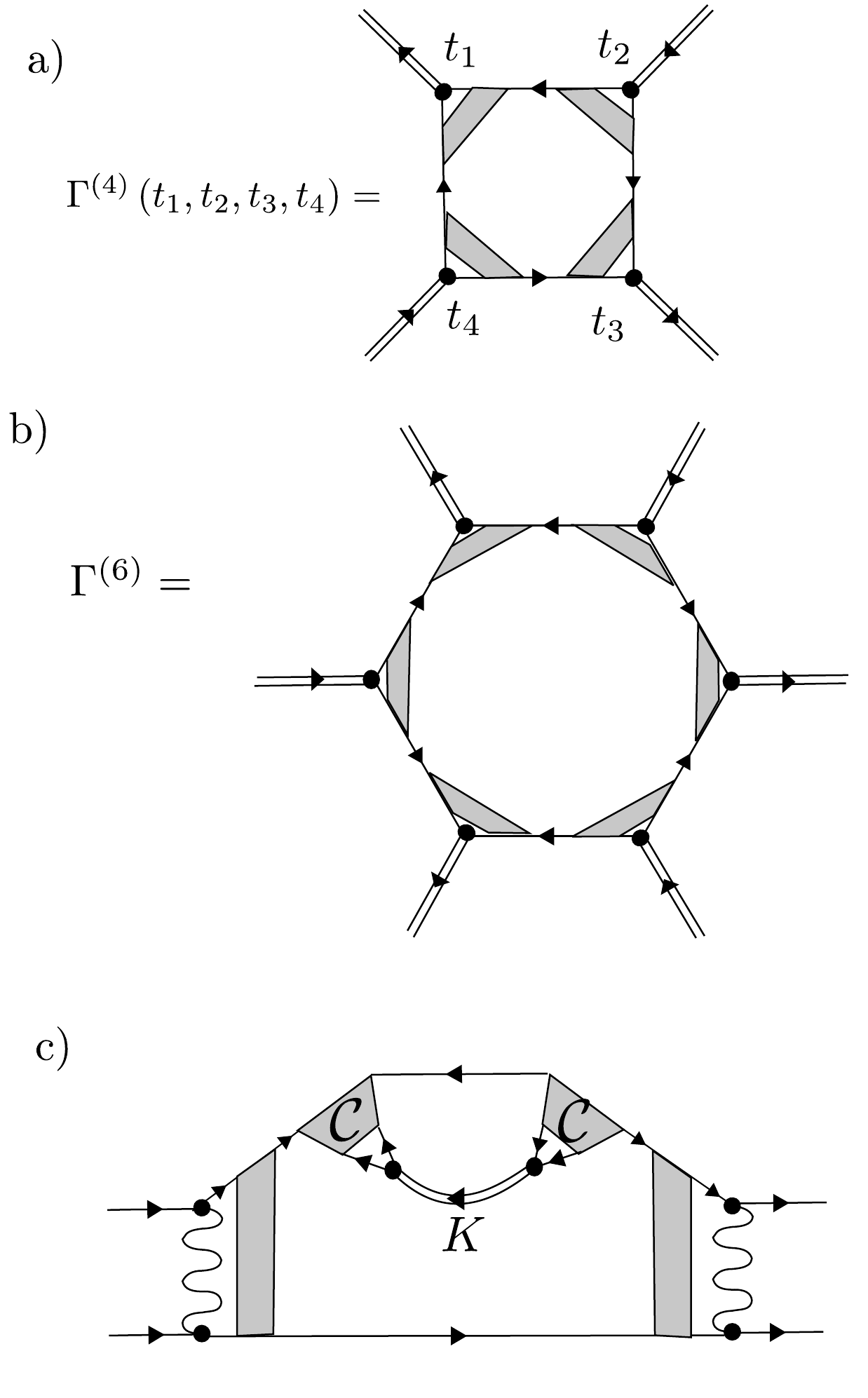}
\caption{ a) Fermion loop with four fluctuation vertices, b) Fermion loop with six vertices. c) Contribution to $\Pi_R$. \label{fig:GLApp}
}
\end{figure}

For fermion loops, following our earlier discussion, these may be treated as "short-ranged objects" where all times are separated on the scale of
at most $\hbar/T$. Thus it may be treated as a delta-function. For example the fermion loop of Fig.~\ref{fig:GLApp}(a) may be estimated as
\begin{equation}
	\Gamma^{(4)}\left(t_1,t_2,t_3,t_4\right) \sim  \frac{\nu}{T^2}\delta(t_1 - t_2)\delta(t_2 - t_3)\delta(t_3-t_4).
\end{equation}	
Above we have used that a fermion loop
involves a momentum integral which gives a factor of the density of states $\nu$. The coefficient then follows on dimensional grounds
$\frac{\nu}{T^2}$ and the fact that $T\tau \ll 1$.
Similarly we expect a fermion loop with $n$ vertex insertions
to be a delta function times $\nu T^{2-n}$.

Since every fluctuation ends in two vertices, we assign  the overall factor $B T/\nu$ for the fluctuation lines as a factor of
$(B T/\nu)^{1/2}$ for each vertex. Further, each fluctuation line brings an integral of momentum $\int d^2 q$ which can be estimated as $\xi^{-2}$. 
Every fermion loop imposes conservation of momentum in the outgoing fluctuation. This should similarly be estimated as $\xi^2$.
Combining all of these estimates we have that the fermion loop with $n$ fluctuation vertices should be estimated as
\begin{align}
 \Gamma^{(n)} &\sim \left(\frac{B T}{ \xi^2\nu}\right)^{n/2}
 \left(
 	\frac{\nu}{T^{n-2}}\right)
 \frac{\xi^2}{B}\nonumber\\\
 	&\sim T \left(\frac{B}{\xi^2 \nu T}\right)^{n/2 - 1},
\end{align}
times the appropriate delta functions. Thus we may neglect diagrams with more than 2 fluctuation vertices in a fermion loop if the quantity $B/(\xi^2 \nu T)$ is small.
Using the fact that $\xi^2 \sim v^2_F\tau / T$, $\nu\sim p_F/v_F$, we see that this gives the criterion $B/E_F\tau \ll 1$

For concreteness we consider the correction to $\Pi_R$ given by  Fig~\ref{fig:GLApp} (c). This is a diagram of type $\Gamma^{(4)}$
with two outgoing fluctuation lines connected, so we may estimate it similarly as
\begin{align}
\frac{1}{\nu}\delta\Pi_R
	&\sim
	\frac{1}{\nu}
	\frac{\nu}{T^2}
	\int
	\frac{d^2 q}{(2\pi)^2}
	D_K\!\left(t,t;q\right)
	\nonumber\\
&\sim
	\frac{1}{\nu T\xi^2}
	\!\int dx x B\left(t, x/\xi\right),
\end{align}
Since this is a correction to the $q$-independent part of $\Pi_R$ we may treat it as a correction to $\epsilon(t)$. Inserting this into the
kinetic equation~\eqref{eq:kineticB}, we obtain,
\begin{align}
&\left[\partial_t  + \frac{16T}{\pi}\left(\epsilon(t) + x^2\right) \right] B(t,x/\xi) = T\biggl(1 \nonumber\\
&+a\frac{B\left(t,x/\xi\right )}{E_F \tau}\int dx' x' B\left(t,x'/\xi\right) \biggr),
\end{align}
for some $\mathcal{O}(1)$ factor $a$. The additional factor is quadratic in $B$. As the integral factor on the RHS is only weakly singular 
$\sim \log \epsilon$ in equilibrium,
it is sufficient to logarithmic accuracy to insist on the stated criterion $B/E_F\tau \ll 1$.

For example we may consider the hard quench where $B$ is small for $t < 0$ and then evolves under some fixed $\epsilon = 0$ for $t >0$.
We first solve the equation neglecting that quadratic term on the RHS, giving
\begin{equation}
B^0(t,x/\xi) = \frac{1-e^{-tx^2}}{x^2},
\end{equation}
where we have set $\pi/16T = 1$. Plugging this into the quadratic term with $q=0$ gives $\sim (t \log t) /E_F\tau$. Thus the quadratic term will be a small correction as long as this is $\ll 1 $, which to logarithmic accuracy is $t \ll E_F\tau$. As $B_{q=0} = t$, this is equivalent to the stated criterion $B/E_F\tau\ll 1$.

Consider another example, a quench $\epsilon(t)  = -\epsilon_f < 0$ for $t>0$ and $\epsilon \sim 1$ for times $t<0$. The solution of
Eq.~\eqref{eq:kineticB} gives that,
\begin{equation}
B^0\left(t,x/\xi\right) = \frac{1-e^{- tx^2 +\epsilon_f t}}{x^2-\epsilon_f}.
\end{equation}
Inserting this into the correction term as before, we may approximate the integral over $x'$ by noting that it is dominated by $x' < 1/\sqrt{t}$.
Thus, the $x'$ dependence in the denominator may be neglected if $\left|\epsilon_f\right| t \gg 1$. This gives that the correction term for $q = 0$ is given by
$\exp(2\epsilon_f t)/\epsilon_f t E_F \tau$. Insisting that this is $\ll 1$ then gives, to logarithmic accuracy, 
the criterion stated in the Introduction, $t \ll \epsilon_f^{-1}\log(E_F\tau)$.

\section{Further details in deriving $\sigma^{\rm AL}$ in Eq.~\eqref{eq:sigAL}}\label{kuboAL}

We now discuss the technical steps involved in obtaining $\sigma^{\rm AL}$. Note that the AL diagram has two fermion loops,
one on the left and one on the right. Let us discuss the left fermion loop first. There are 7 distinct diagrams that
contribute to the left loop. Of these three are straightforward to guess. Basically it is the left loop in
Fig.~\ref{fig:feyn} (i) but with Keldysh propagator corresponding to three different locations. Going clockwise along
the times $s_1\rightarrow t_1\rightarrow s_3\rightarrow s_1$, these 3 diagrams are 1). $G_RG_KG_A$ (shown in the figure). 2). $G_R G_A G_K$
and 3).  $G_KG_AG_R$. The Cooperons appear in all these diagrams the same way.

We now account for the remaining 4 diagrams. Impurity lines can only connect $G_R$ and $G_A$.
This is clear from the structure of the Cooperons.
However, since $G_K$ contains both $G_R$ and $G_A$, we need to also account for
impurity lines connecting $G_R,G_K$ as well. We also have to account for impurity lines connecting $G_A,G_K$.

Thus the remaining 4 diagrams are accounted for by dropping an impurity
line between a $G_K,G_R$ propagator or between a $G_K,G_A$ propagator. For diagram Fig.~\ref{fig:feyn} i). one can only drop an additional
impurity line between $G_K$ and $G_A$ giving a diagram where the Green's functions
appear as (leaving out the Cooperon locations below, and going clockwise from $s_1\rightarrow t_1\rightarrow s_3\rightarrow s_1$)
\begin{align}
{4).}\,\, &G_R\,{\rm current-vertex}\, G_A\, {\rm impurity-vertex}\, G_K \nonumber\\
&G_A\,{ \rm impurity-vertex}\, G_A.\nonumber
\end{align}
For diagram 2) one proceeds similarly to obtain diagram 5).
For diagram 3) there are two possible ways to drop an impurity line since
one can drop an impurity line between $G_K$ and both $G_R$ and $G_A$. This gives the last two of the set of
seven.

Dropping these additional impurity lines
is explicitly shown for the Maki-Thompson diagrams in Fig.~\ref{fig:feyn} j) and k). In Matsubara notation, only Fig.~\ref{fig:feyn} j) arises.
But in Keldysh notation, the book-keeping requires keeping track of this additional kind of diagrams such as Fig.~\ref{fig:feyn} k).
Note that if all Green's functions were the same (as it is in the Matsubara notation), then Fig.~\ref{fig:feyn} k) is identical to Fig.~\ref{fig:feyn} j)
as the extra impurity line can be absorbed into the definition of the Cooperons.
The next sub-section explicitly carries out a computation of such terms in the context of the MT conductivity.

Now we discuss the right loop. There are a total of 4 possible combinations. One of them is shown in the figure and
corresponds to, going from $s_2\rightarrow s_4\rightarrow t_2 \rightarrow s_2$, 1): $G_RG_KG_R$.  It is clear from the structure of the Cooperons
that the impurity lines only connect $G_R$ and $G_A$. That is why for this diagram, the Cooperon only appears on the top
between $G_R,G_K$ Green's functions. Going along the same direction, diagram 2) is $G_RG_AG_K$, where $G_K$ now enters between times $s_2,t_2$.
In this case, the right loop indeed has two Cooperons and looks like the mirror reflection of the left loop.  Diagram 3) is
now the extension of diagram 1) where an additional impurity line is dropped between the $G_K$ propagator and the $G_R$
propagator on the opposite side. While diagram 4) is a similar extension of diagram 2).

Since the insertion of the Cooperon implies adding a term like $\mathcal{P}_q \equiv 1/(2\pi\nu\tau)\sum_k G_R(k) G_{A/K}(k)= {\mathcal O}(1)$,
all the diagrams listed above with their different combinations of Cooperons will contribute equally.

Now we discuss the calculation of one of these diagrams.
The transient optical conductivity calculation relies on an important separation of time-scales.
Note that the electron Green's functions in real time behave as,
\begin{align}
G_{R}(t,t',k) = -i\theta(t-t') e^{-i\epsilon_k (t-t')-(t-t')/2\tau}.
\end{align}
Thus the  electron Green's function is exponentially suppressed when $t-t'>\tau$. Since, $T\tau\ll 1$, this implies
that the Fermion loops in any diagram have significant support only within a time window $t-t' < \tau, T^{-1}$. Since the
superconducting fluctuations are governed by a much longer time-scale, when diagrams involve both fermion loops and boson-loops,
the former can be treated as delta functions in time relative to the latter.

Thus the left fermion loop of Fig.~\ref{fig:feyn} (i) may be approximated as
a delta-function $\delta(t_1-s_3)\delta(t_1-s_1)$. This is equivalent to an integration over all frequencies of the
form,
\begin{align}
{\rm left-loop} \equiv &\int \frac{d\omega}{2\pi}\sum_k v_k^i\left[\mathbb{C}_q(\omega,-\omega)\right]^2i G_R(k,\omega)\nonumber\\
&\times i G_A(-k+q,-\omega)i G_K(k,\omega).
\end{align}
Above, $v_k^i$ is the Fermion velocity arising due to
the current vertex.
Similar approximations lead to the final Equations.~\eqref{eq:sigAL}, where the left and right loops each contribute
a factor proportional to $\nu \xi^2 q$.  Finally the fluctuation part appears as $D_K(t_1,t_2)D_R(t_1,t_2)$, which using the
definition of $D_K$ in terms of $B_q$, namely $D_K \equiv 4 D_R B_q$ reduces to Eq.~\eqref{eq:sigAL}.

These also reduce to the equilibrium results when the detuning $\epsilon$
is constant in time.

\section{Further details in deriving $\sigma^{\rm MT}$ in Eq.~\eqref{eq:sigMT}}\label{kuboMT}

Let us first consider Fig.~\ref{fig:feyn} (j). Consider the region of the diagram marked by times $s_{1,2,3,4}$ and
denote it by ${\rm MT}_q(s_1,s_2,s_4,s_3)$. Then, this diagram has the form,
\begin{align}
{\rm MT}_q(s_1,s_2,s_4,s_3) &=\int ds \int ds'  {\mathbb C}_q(s_1,s,s_2,s) iD_K(q,s,s') \nonumber\\
&\times {\mathbb C}_q(s',s_4,s',s_3).
\end{align}
Using Eq.~\eqref{eq:longtimeC}, the time -integrals $s,s'$ can be easily performed to give,
\begin{align}
&{\rm MT}_q(s_1,s_2,s_4,s_3) = \frac{1}{4\tau^2}\theta(s_1-s_2)\theta(s_3-s_4) \nonumber\\
&\times e^{-\frac{1}{2}(s_1-s_2)(D q^2+1/\tau_{\phi})}e^{-\frac{1}{2}(s_3-s_4)(D q^2+1/\tau_{\phi})}\nonumber\\
&\times iD_K\biggl(q,\frac{s_1+s_2}{2}, \frac{s_3+s_4}{2}\biggr).
\end{align}

Thus the diagram becomes,
\begin{align}
&{\rm 1(j)} = \frac{1}{2}\sum_q \int_{s_{1,2,3,4}}\sum_kv_k^i v^j_{-k+q}{\rm MT}_q(s_1,s_2,s_4,s_3)\nonumber\\
&\times iG_R(k,t_1,s_3)iG_K(-k+q,t_2,s_4)\nonumber\\
&\times iG_A(k,s_1,t_1)iG_R(-k+q,s_2,t_2). \label{mtd1}
\end{align}

Since as compared to ${\rm MT}_q$, the Green's functions are rapidly varying as they are
peaked at times short as compared to $\tau,1/T$, we make the following approximations for
the retarded electron Green's function:
\begin{align}
&G_R(k,t) = \int \frac{d\omega}{2\pi}\frac{e^{-i\omega t}}{\omega -\xi_k +i/2\tau} \nonumber\\
&\simeq \int \frac{d\omega}{2\pi}\frac{e^{-i\omega t}}{-\xi_k +i/2\tau} = \frac{\delta(t)}{-\xi_k+i/2\tau}\theta(t).
\end{align}
A similar approximation for the Keldysh electron Green's function gives,
\begin{align}
G_K(k,t) &= -i\tau^{-1}\int \frac{d\omega}{2\pi}\frac{e^{-i\omega t}n(\omega)}{(\omega -\xi_k)^2 +(1/2\tau)^2} \nonumber\\
&\simeq  -i\frac{1}{2T\tau}\int \frac{d\omega}{2\pi}\omega\frac{e^{-i\omega t}}{(\omega-\xi_k)^2 +(1/2\tau)^2} \nonumber\\
&= \frac{1}{2T\tau}\biggl[\frac{\partial_t\delta(t)}{\xi^2_k+(1/2\tau)^2}\biggr].
\end{align}

In Eq.~\eqref{mtd1} we may replace $v_k^{i}v_{-k+q}^j\simeq -v_k^i v_k^j$, and evaluate all the Green's functions at $q=0$, but keep
the $q$ dependence in ${\rm MT}_q$. Changing the sum on $k$ into an integral over energies, and noting that angular integral of
$v^i_k v^j_k$ over the Fermi surface equals $\delta^{ij}v_F^2/2$, we obtain,
\begin{align}
&{\rm 1(j)} = -\frac{1}{4}\sum_q v_F^2\int \nu d\xi \int_{s_{4}} {\rm MT}_q(t_1,t_2,s_4,t_1)\nonumber\\
&\times \left[\frac{1}{\xi^2+(1/2\tau)^2}\right]^2\frac{1}{-\xi+i/2\tau}\frac{1}{2T\tau}\partial_{s_4}\delta(s_4-t_2).
\end{align}
After integration over energy, and also an integration by parts in $s_4$, the above reduces to,
\begin{align}
&{\rm 1(j)} = -\frac{1}{8}\sum_q v_F^2 \frac{\nu3i\pi \tau^3}{T}\partial_{s_4}{\rm MT}_q(t_1,t_2,s_4,t_1)\biggl|_{s_4=t_2}.
\end{align}
Above we have accounted for the fact that integral over delta function times step function gives a $1/2$.

Now we turn to the diagram Fig.~\ref{fig:feyn}(k). This is different from the above mentioned figure by the addition
of another impurity line. While in the Matsubara formalism, there is only one kind of Green's function, this diagram is topologically
the same as Fig.~\ref{fig:feyn}(j). However, in the Keldysh formalism, this additional impurity line inserts a $G_KG_A/(2\pi\nu\tau)$
combination to the usual $G_RG_A/(2\pi\nu\tau)$ series accounted for in the Cooperon diagrams.
Thus this diagram takes the form,
\begin{align}
&{\rm 1(k)} = \frac{1}{2}\sum_q \int_{s_{1,2,3,4}}\sum_kv_k^i v_{-k+q}^j{\rm MT}_q(s_1,s_2,s_4,s_3)\nonumber\\
&\times iG_R(k,t_1,s_5)iG_R(-k+q,t_2,s_6)\nonumber\\
&\times iG_A(k,s_1,t_1)iG_R(-k+q,s_2,t_2)\nonumber\\
&\times \frac{-1}{2\pi\nu\tau}\sum_{k'}\int_{s_{5,6}}iG_R(k',s_5,s_3)iG_K(-k'+q,s_6,s_4).\label{mtd2}
\end{align}
Making the same approximations as before for the Green's functions, we obtain,
\begin{align}
&{\rm 1(k)} = -\frac{1}{4}\sum_qv_F^2 \int ds_4 {\rm MT}_q(t_1,t_2,s_4,t_1)\nonumber\\
&\times \nu\int d\xi \frac{1}{\xi^2+(1/2\tau)^2}\frac{1}{(-\xi+i/2\tau)^2}\nonumber\\
&\times \frac{-\nu}{2\pi\nu\tau}\int d\xi' \frac{1}{-\xi'+i/2\tau} \frac{1}{\xi'^2+(1/2\tau)^2}\frac{1}{2T\tau}\partial_{s_4}\delta(s_4-t_2).
\end{align}
The above simplifies to,
\begin{align}
&{\rm 1(k)} = -\frac{1}{8}\sum_qv_F^2 \frac{\nu \pi i \tau^3}{T}\partial_{s_4}{\rm MT}_q(t_1,t_2,s_4,t_1)\biggl|_{s_4=t_2}
\end{align}
Combining Fig.~\ref{fig:feyn} (j,k) we obtain,
\begin{align}
{\rm 1(j)+1(k)}= -\frac{1}{8}\sum_qv_F^2 \frac{\nu 4 \pi i \tau^3}{T}\partial_{s_4}{\rm MT}_q(t_1,t_2,s_4,t_1)\biggl|_{s_4=t_2}.
\end{align}
There are two more diagrams (not shown, lets call them 1 j',k'). These are obtained from the above two diagrams by
a). interchanging $G_R$ and $G_A$, b). taking $G_K\rightarrow -G_K$, c). interchanging $s_1,s_3$, d). interchanging
$s_2,s_4$.
Then the total MT contribution is,
\begin{align}
{\rm 1(j,k,j',k')} &=   -\frac{1}{8}\sum_qv_F^2 \frac{\nu 4 \pi i \tau^3}{T}
\biggl[\partial_{s_4}{\rm MT}_q(t_1,t_2,s_4,t_1)\biggl|_{s_4=t_2}\nonumber\\
&+\partial_{s_4}{\rm MT}_q(t_1,s_4,t_2,t_1)\biggl|_{s_4=t_2}\biggr]\nonumber\\
&= -\frac{1}{8}\sum_qv_F^2 \frac{\nu 4 \pi i \tau^3}{T} \partial_{t_2}\biggl[{\rm MT}_q(t_1,t_2,t_2,t_1)\biggr].
\end{align}
The above is a current-current correlation function.
The current due to MT terms is obtained from integrating the above with the vector potential $A(t)$. This gives
\begin{align}
J^{\rm MT}(t_1)&=  i \int_{-\infty}^{t_1} dt_2  \left(-\frac{1}{8}\right)
\sum_qv_F^2 \frac{\nu 4 \pi i \tau^3}{T} \nonumber\\
&\partial_{t_2}\biggl[{\rm MT}_q(t_1,t_2,t_2,t_1)\biggr]
A(t_2).
\end{align}
Integrating by parts, and noting that $\partial_t A(t) = -E(t)$, and using the definition of the diffusion constant,
\begin{align}
\sigma^{\rm MT}(t_1,t_2)= \frac{D\pi \tau^2}{T}{\rm MT}_q(t_1,t_2,t_2,t_1).
\end{align}
Plugging in that $iD_K(t,t) = \left(32 T/\pi\nu\right) B_q(t)$, we obtain Eq.~\eqref{eq:sigMT}.

The regular MT term in Fig.~\ref{fig:feyn} (l) is small as compared to the terms we just evaluated because the Keldysh
Green's function in this diagram is $G_K(t_1,s_1)\sim \partial_{s_1}\delta(t_1-s_1)$, where $t_1$ is the largest time. 
Performing the same manipulations as above, we find that the 
current measured at time $t_1$ comes from short lived processes.  

\section{Equivalence to equilibrium} \label{condeq}
In equilibrium the conductivity may be equivalently expressed in terms of the Fourier transform. In equilibrium, $B_q(t)$ is of course
time independent and thus Eq.~\eqref{eq:defDR} and Eq.~\eqref{eq:kineticB} imply that
\begin{gather}
D_R(t,t') = \frac{8T}{\pi\nu}
\exp\left[-\frac{8T}{\pi}(t-t')
	\left(
		\epsilon + \xi^2q^2
	\right)
	\right]\theta(t-t');
\\
B_q = \frac{\pi}{16}\frac{1}{\epsilon +\xi^2 q^2}.
\end{gather}
If we substitute this into the formula for $\sigma^{\rm AL}(t,t')$ given in Eq.~\eqref{eq:sigAL}, and Fourier transform, we obtain,
(defining $\bar{\omega}\equiv \frac{\pi\omega}{16 T}$),
\begin{widetext}
\begin{align}
&\sigma^{\text{AL}}(\omega)=
	\int_{-\infty}^\infty dt'
	e^{i\omega(t-t')} \sigma(t,t')=
32\int dt'e^{i\omega(t-t')}\int_{-\infty}^{t'}\!\!\ ds\int\! \frac{d^2 q}{(2\pi)^2}\, \xi^4 q^2 \nu^2|D_R(q,t-s)|^2 B_q
\nonumber\\
&=
2\pi\int dt'e^{i\omega(t-t')}
\int_{-\infty}^{t'}\!\!\ ds
\int\! \frac{d^2 q}{(2\pi)^2}\, \xi^4 q^2\frac{
	\left(
		\frac{8T}{\pi}
		e^{-\frac{8T}{\pi}(t-s)(\epsilon + \xi^2q^2)}
	\right)^2
	}{
	\epsilon + \xi^2q^2
}\nonumber\\
&= 2\pi\int dt'e^{i\omega(t-t')}
\int\! \frac{d^2 q}{(2\pi)^2}\, \xi^4 q^2 \
(\frac{8T}{\pi})^2\frac{\pi}{16T}\frac{
		e^{-\frac{16T}{\pi}(t-t')(\epsilon + \xi^2q^2)}
	}{
	\left(\epsilon + \xi^2 q^2\right)^2
}
\nonumber\\
&=
8T \int_0^{\infty}\frac{dx}{4\pi}
\frac{x}{\left(x+\epsilon\right)^2}
\int dt'
e^{
	-\frac{16T}{\pi}(t-t')\left(-i\bar{\omega} +x+ \epsilon
	\right)}=
\frac{1}{8}\int_0^\infty dx
\frac{x}{(x+\epsilon)^2(-i\bar{\omega} + x+ \epsilon)}
\nonumber\\
&=
\frac{1}{16\epsilon}
	\frac{2\epsilon^2}{\bar{\omega}^2}
	\bigg(
		-\frac{ i\bar{\omega}}{\epsilon}
		- (1-\frac{i\bar{\omega}}{\epsilon})
		\log(1 - \frac{i\bar{\omega}}{\epsilon} )
	\bigg).
\end{align}
\end{widetext}

Similarly for $\sigma^{\rm MT}$ given in Eq.~\eqref{eq:sigMT}, we have,(defining $\gamma_\phi \equiv  \frac{\pi}{8T\tau_\phi}$),
\begin{widetext}
\begin{align}
&\sigma^{\rm MT}(\omega)
=
	\int_{-\infty}^\infty dt'
	e^{i\omega(t-t')}
	\sigma(t,t')
\nonumber\\
&= \int dt'e^{i\omega(t-t')}
	8 D \!\! \int\!\!\!
	\frac{d^2q}{(2\pi)^2}
	B_q
	e^{
		-\left(\frac{1}{\tau_\phi} + D q^2\right)
		\left( t-t'\right)
	}
= 8 D \!\! \int\!\!\!
	\frac{d^2q}{(2\pi)^2}
	\frac{1
	}{
		-i\omega
		+ \left(
			\frac{1}{\tau_\phi}
			+ D q^2
		\right)
	}
	\frac{\pi}{16}
	\frac{1}{\epsilon + \xi^2q^2}
\nonumber\\
&=
\frac{\pi}{2}
\int_0^{\infty}
\frac{dx}{4\pi}
\frac{1
	}{-i
	\frac{\pi \omega}{8 T}
	+
	\gamma_\phi + x}
\frac{1}{\epsilon + x}= \frac{1}{8\epsilon}
	\left(
		\frac{-2i\bar{\omega} +\gamma_\phi}{\epsilon}
		- 1
	\right)^{-1}\!\!\!
	\log\left(
		\frac{-2i
			\bar{\omega} +\gamma_\phi
		}{
			\epsilon
		}
	\right).
\end{align}
\end{widetext}
These may be compared with the results in for example Ref.~\onlinecite{Federici97}.
This gives asymptotically in the various limits that
\begin{widetext}
\begin{equation}
\sigma^{\rm MT}\!\left(\bar{\omega}\right) + \sigma^{\rm AL}\!\left(\bar{\omega}\right) = \frac{1}{16\epsilon}
\begin{cases}
\frac{3\epsilon}{-i\bar{\omega}} \log\frac{-i\bar{\omega}}{\epsilon}
\qquad
&\bar{\omega} \gg \epsilon,\gamma_\phi
\\
 1 - 2\log \frac{-i\bar{\omega}}{\epsilon}
\qquad
& \epsilon \gg \bar{\omega} \gg \gamma_\phi
\\
\frac{2\epsilon}{-i \bar{\omega}}\log \frac{-i\bar{\omega}}{\epsilon} + \frac{2\epsilon}{\gamma_\phi} \log\frac{\gamma_\phi}{\epsilon}
\qquad
&
\gamma_\phi\gg \bar{\omega} \gg \epsilon
\\
1 +2\frac{\epsilon}{\gamma_\phi- \epsilon}\log\frac{\gamma_\phi}
{\epsilon}
\qquad
 &\epsilon,\gamma_\phi \gg \bar{\omega}
\end{cases}.
\end{equation}
\end{widetext}

For example, if $\epsilon \gg \gamma_\phi$, we have that from $1 \gg \bar{\omega} \gg \epsilon$ the real part of $\sigma$ goes as $1/|\bar\omega|$.
When $\bar{\omega} \sim \epsilon$
this crosses over to a weaker logarithmic growth with $\bar\omega$, finally saturating to
$\frac{1}{8\epsilon}\log \epsilon/\gamma_\phi$ at $\bar \omega \sim \gamma_\phi$.

\section{Experimental definition of $\sigma(\omega)$}      \label{sigE}

\begin{figure}
\includegraphics[width = 3.5in]{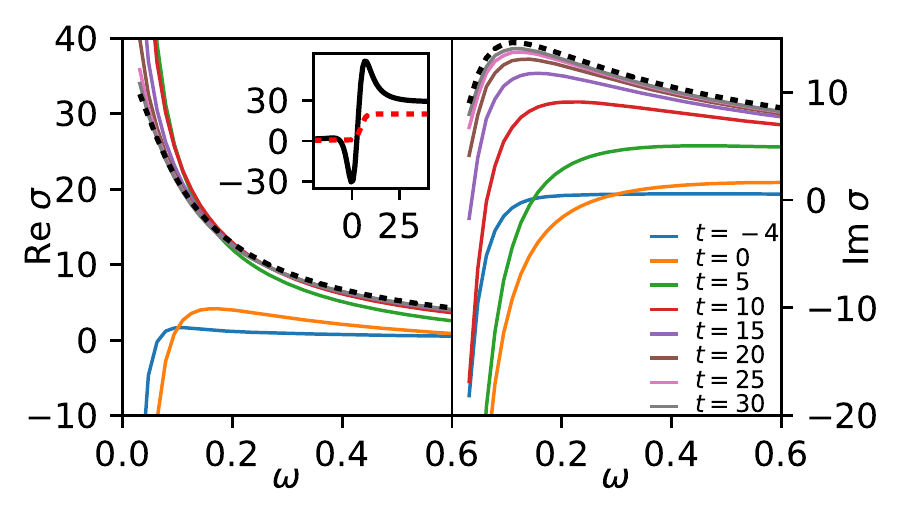}
\caption{The "experimental" fluctuation correction to the optical conductivity shown for several different times under the detuning given in Fig.~\ref{fig:flucs},
left panel. All times and inverse frequencies given in units of $\pi\hbar/8T$. Inset: Re$\sigma$ at $\pi\omega/8T=.05$,
showing the non-monotonic behavior of the "experimental" conductivity. The red dashed line is the detuning $\epsilon(t)$ shown at arbitrary scale.
\label{fig:expt_saturation}
}
\end{figure}

\begin{figure}
\includegraphics[width = 3.5in]{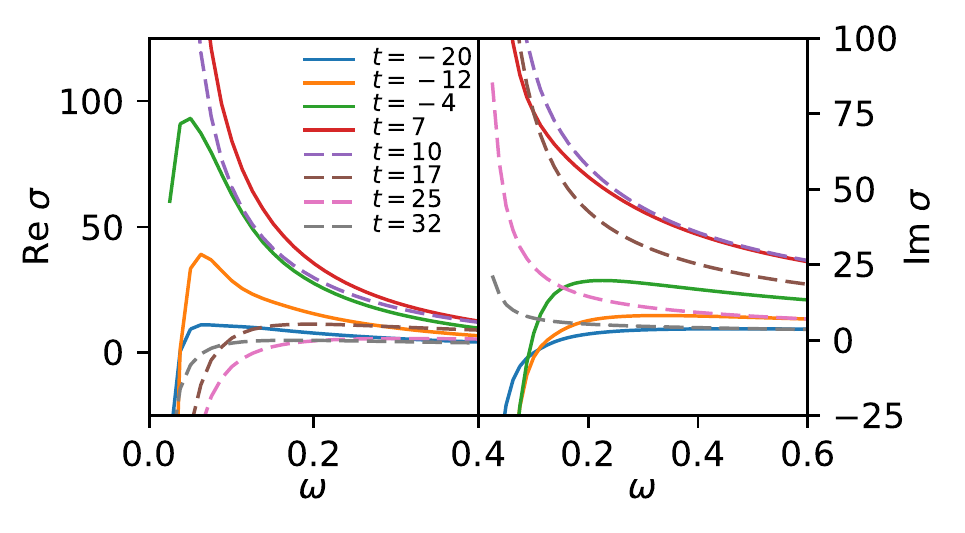}
\caption{The "experimental" fluctuation correction to the optical conductivity shown for several different times under the detuning given in Fig.~\ref{fig:flucs},
right panel.
All times and inverse frequencies in units of $\pi\hbar/8T$.
\label{fig:expt_smoothquench}
}
\end{figure}

In this section we consider an electric field pulse $E(t) = f(t-t_0)$, centered at $t_0$, and calculate the induced current $J(t)$.
We then Fourier transform $J$, $E$ and compute the experimental conductivity "$\sigma(\omega,t_0)$" $= J(\omega)/E(\omega)$.
This experimental conductivity is plotted in Figs.~\ref{fig:expt_saturation},~\ref{fig:expt_smoothquench}. We consider here the pulse shape
$f(t) = (1-2\frac{t^2}{w^2})\exp( - \frac{t^2}{w^2})$. We set $w = 2 (\pi/ 8T)$ although we do not find particular
sensitivity to this parameter.

The primary difference between the conductivity shown here and the conductivity discussed in the main text, is that the "experimental"
definition leads to highly
non-monotonic behavior of the conductivity at points when the system is far from adiabaticity see Fig.~\ref{fig:expt_saturation}, inset.

This wild behavior at low frequencies can be understood as follows. The Fourier transform $E(\omega)$ goes to zero as $\omega \rightarrow 0$ for
experimentally feasible pulses.
As the quantity we are considering is $J(\omega)/E(\omega)$, the vanishing of the denominator would be problematic. In steady state of course this is not a problem because
if $E(\omega)$ vanishes then $J(\omega)$ must also vanish and so the ratio is finite. However, in the "experimental" calculation, when there is no time-independence,
there is no reason why $J(\omega\rightarrow 0)$ may not be finite, and thus the ratio $J/E$ will be singular as $\omega\rightarrow 0$.
This implies that there will be large differences between the different notions of conductivity at low frequencies when the system is farthest from adiabaticity.
This appears as an overshooting behavior, and is most apparent in Fig.~\ref{fig:expt_saturation} where many solid lines
over-shoot the equilibrium value (black dashed lines), with the over-shooting the largest at low frequencies.

\end{appendix}


%

\end{document}